\documentclass[conference]{IEEEtran}
\IEEEoverridecommandlockouts
\usepackage{cite}
\usepackage{amsmath,amssymb,amsfonts}
\usepackage{algorithmic}
\usepackage{graphicx}
\usepackage{textcomp}
\usepackage{xcolor}
\usepackage[textsize=tiny]{todonotes}
\usepackage[acronym]{glossaries}
\usepackage{subcaption}
\usepackage{tikz}
\usepackage{pgf-pie}
\usepackage{enumitem}
\usepackage{multirow}

\usepackage{pgfplots}
\pgfplotsset{width=7cm,compat=1.8}

\def\BibTeX{{\rm B\kern-.05em{\sc i\kern-.025em b}\kern-.08em
    T\kern-.1667em\lower.7ex\hbox{E}\kern-.125emX}}

\newacronym{eu}{EU}{European Union}
\newacronym{pv}{PV}{Photo Voltaic}
\newacronym{ttp}{TTP}{Trusted Third Party}
\newacronym{pos}{PoS}{Proof-of-Stake}
\newacronym{post}{PoST}{Proof-of-StakeTime}
\newacronym{aura}{AuRa}{Proof-of-Authority}
\newacronym{lpd}{LPD}{Local Power Distributor}
\newacronym{dso}{DSO}{Distribution System Operator}
\newacronym{ev}{EV}{Electric Vehicle}
\newacronym{hw}{HW}{Hardware}
\newacronym{sw}{SW}{Software}
\newacronym{lora}{LORA}{Long Range LPWAN}
\newacronym{gprs}{GPRS}{General Packet Radio Services}
\newacronym{plc}{PLC}{Power-Line Communication}
\newacronym{p2p}{P2P}{Peer-to-Peer}
\newacronym{blc}{BLC}{Blockchain}
\newacronym{dlt}{DLT}{Distributed Ledger Technologies}
\newacronym{pow}{PoW}{Proof-of-Work}
\newacronym{poa}{PoA}{Proof-of-Authority}

\usepackage[hyphens]{url}

\begin{document}

\title{Towards a Peer-to-Peer Energy Market: an Overview
%Future Energy Market: a Short Status and Outlook Analysis towards Decentralised Power Production. 
%Requirements and Lacks of Electrical Power Grid Towards a Decentralised Power Production\\
% {\footnotesize \textsuperscript{*}Note: Sub-titles are not captured in Xplore and
% should not be used}
%\thanks{The work leading to this paper is partially financed by the European Commission, thanks to the EU Blockchain Observatory \& Forum. This is an initiative sponsored by the European Commission, Directorate-General of Communications Networks, Content \& Technology.}
}
\author{
\IEEEauthorblockN{Luca Mazzola, Alexander Denzler and Ramón Christen}
\IEEEauthorblockA{\textit{Information Technology} \\
\textit{HSLU - Lucerne University of Applied Sciences and Arts}\\
Suurstoffi 1, 6343 Risch-Rotkreuz, Switzerland \\
\{luca.mazzola, alexander.denzler, ramon.christen\}@hslu.ch}
}

\maketitle

%\todo{@Alex: should we already change the authors order?}

\begin{abstract}
%This paper provides an overview of the current state of the energy market, with respect to the increasing number of decentralised prosumers. 
%Beginning with outlining the limitations imposed by the status quo, a potential multi-layered architecture of a \gls{p2p} energy market will be introduced.  This paper will then discuss the fundamental aspects of local production and local consumption as part of a microgrid along with changes in roles and some incentive models connected with decentralisation. 
%To give the reader a feeling of it, a typical \gls{p2p} settlement is explained.
%Also relevant elements of trading, such as Smart contract and grid stability are presented to the reader. 
%A review of relevant activities is the presented, to showcase where existing project in this domain are going and what are the most important themes.
%Being this a work in progress, many open questions are still on the table and will be addressed in the next stages of the research. 
%The outcome of this work is firstly to present these open issues, and secondly to provide a reference model that we would like to use as base for further discussions and improvement, in a dialog with the broad community oriented towards a more fair and ecological-friendly solution for the electricity market of the future.
%
This work focuses on the electric power market, comparing the status quo with the recent trend towards the increase in distributed self-generation capabilities by prosumers.
Starting from the existing tension between the intrinsically hierarchical current structure of the electricity distribution network and the substantially distributed and self-organising nature of the self-generation, we explore the limitations imposed by the current conditions.
Initially, we introduce a potential multi-layered architecture for a \gls{p2p} energy  market, discussing  the  fundamental aspects  of  local  production  and  local  consumption  as  part  of  a microgrid.
Secondly, we analyse the consequent changes for the different users' roles, also in connection with some incentive models connected with the decentralisation of the power production.
To give a full picture to  the  reader, we also scrutinise relevant  elements  of energy trading, such as Smart Contract and grid stability. 
Thirdly, we present an example of a typical \gls{p2p} settlement, showcasing the role of all the previously analysed aspects.
To conclude, we performed a  review  of  relevant  activities  in  this  domain, to  showcase  where  existing  projects are  going and  what  are  the  most  important  themes covered.
Being  this  a  work in  progress,  many  open  questions  are  still  on  the  table  and  will be addressed in the next stages of the research. 
Eventually, by providing a reference model as base for further discussions and improvements, we would like to engage ourselves in a dialog with the different users and the broad community, oriented towards a more fair and ecological-friendly solution  for  the  electricity  market  of  the  future.
\end{abstract}


% I: simple intro with research question
% II: status quo vs. 'future' - components and graphic (roles need to be described here)
% III: focus on details -> differences in implemented 'pilot' projects (quartierstrom)
% IV: decentralised system
% A: grid stability - introduction clarifying the two sections: availability and utilities -> we only concentrate on power but not on assets -> we assume it will never be critical -> we only concentrate on the energy flowing in the network (prosumer plus) [needs to be before business because of its incentives]
% B: business (incentives for all the involved roles) 
% C: technology
% V: further research
% VI: conclusion

\section{Introduction}
One of the recent trends in the electric market is the focus on energy self-production from small plants, which are directly located on the consumers premises. 
These installations mainly consist of solar panels installed on  rooves, but have recently included other renewable sources such as small dams or wind turbines.
%\todo{Ramon: add the two introductory aspects, as for the next comments}
%
% start with the undergoing digitalisation of energy systems -> see Andoni intro
% mention what blockchain is and that it allows for P2P networks -> reference see Andoni intro
% volatile power production...
%
  New technologies and decreasing costs in sustainable power production and storage utilities make a local individual-use power production increasingly attractive. 
  As a result, an increasing number of actors using  micro-energy generation has contributed towards the aggregated power production.
  This leads to a more decentralised production, which stands in direct contrast to the legacy production modality, where a small number of centralised producer with a large energy generation have to fully cover the energy demand. 
  The decentralised production that mainly consists of renewable energies sources such as \gls{pv} and wind, has increased within the \gls{eu} from approximately 15\% in 2005 to almost 31\% in 2017~\cite{european_court_of_auditors_wind_2019}. 
  It is noteworthy that concurrent with the increase in total production, the number of different users with renewable energy production possibilities also grew steadily. This indicates a trend towards a distributed self-production. For example, in 2016 Germany registered about 1 million and UK about half a million end users with electricity self-production capabilities from renewable sources~\cite{inderberg2018there}.
  
  This increase is a given in the current power market setting, which does not account for a significant number of self-producing users and not provide any incentive for their further expansions. 
	Strict regulation limits the use of individually produced power: either it is fed-in to the public power grid for a given price set by the \gls{lpd} or it is consumed by the producer
  This issue is being tackled by an ongoing deregulation effort of the energy market, which has opened up new opportunities for self-generated power usage for both  private and industrial parties. 
  One of the aims of market deregulation is to allow customers to freely select the energy provider, which offers the possibility to sell and trade power between any actor. 
  In combination with the undergoing digitisation in power measurement and payment possibilities, this energy market deregulation provides incentives for sharing energy within peer-to-peer energy communities.
  These new incentives, paired with the improvements in capabilities from the technological side, enforce a redesign of the existing business models on all levels. 
  Nevertheless, all these new opportunities bring also new challenges - on the technical, legal as well as the economic side.
  .

  \begin{figure*}[t]
    \centering
    \begin{tabular}[\textwidth]{c|c}
        \centering
        \multirow{4}{*}{
        \begin{subfigure}[t]{0.65\textwidth}
        \vspace{-1.6cm}
            \includegraphics[width=\textwidth]{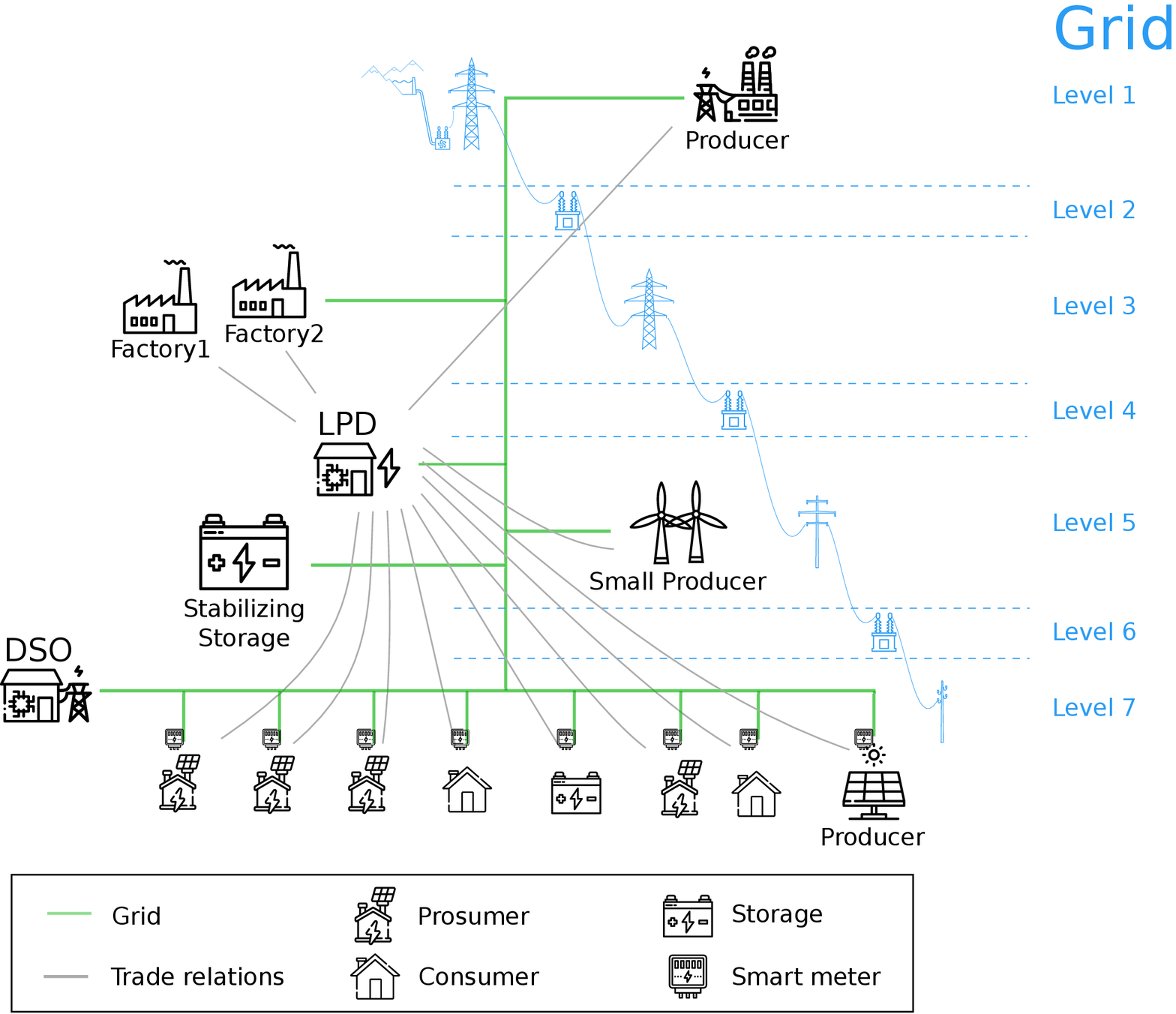}
            \caption{Current star-shaped power grid topology. The power is unidirectional distributed from the producer (small numbers of centralised power plants) to the end customers, through the local power distributor.}
            \label{subfig:current_power_top}
        \end{subfigure}
        }&
        \begin{subfigure}[]{0.3\textwidth}
                \includegraphics[width=\textwidth]{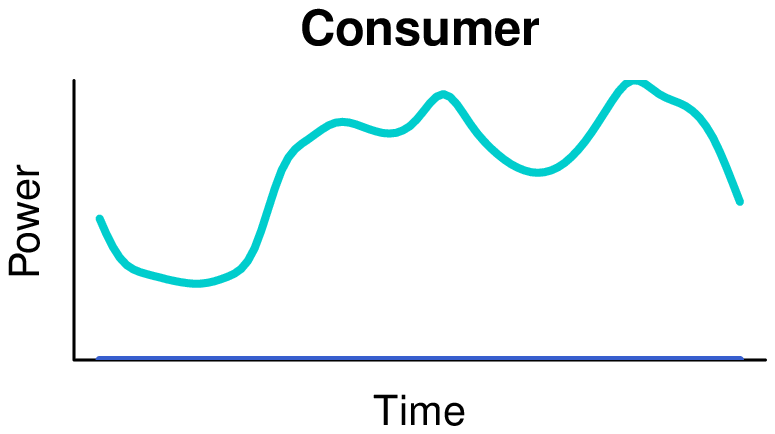}
        \end{subfigure}\\
        &
        \begin{subfigure}[]{0.3\textwidth}
                \includegraphics[width=\textwidth]{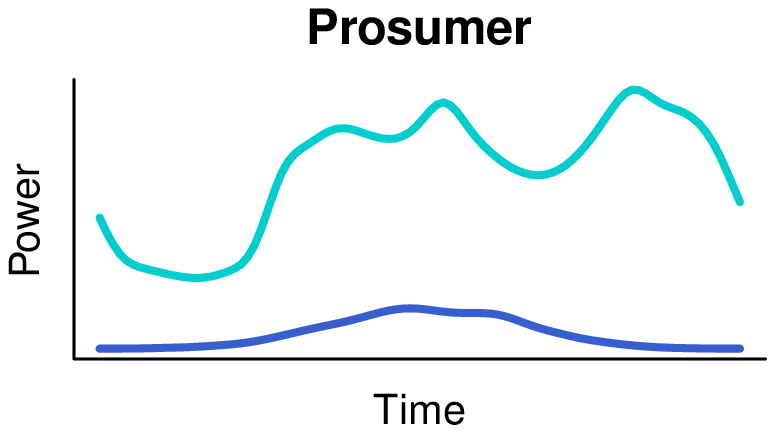}
        \end{subfigure}\\
        &
        \begin{subfigure}[]{0.3\textwidth}
                \includegraphics[width=\textwidth]{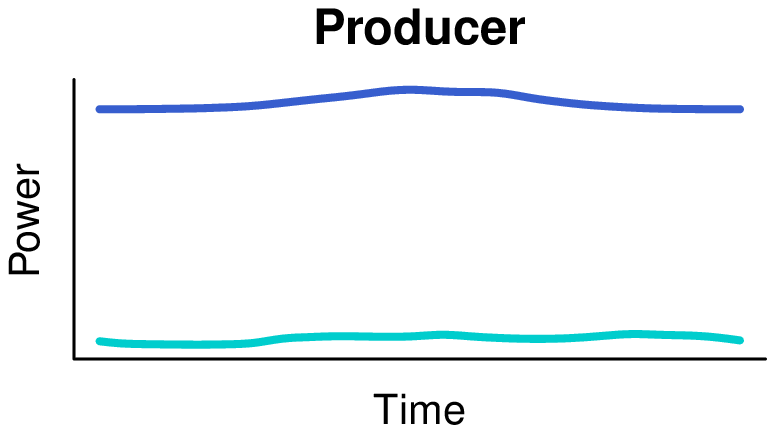}
        \end{subfigure}\\
        &
        \begin{subfigure}[]{0.3\textwidth}
                \includegraphics[width=\textwidth]{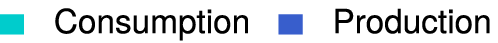}
                \caption{Energy profiles for actors in the current star-shaped power grid set-up.}
                \label{subfig:current_grid_parties}
                \vspace{0.6cm}
        \end{subfigure}\\
    \end{tabular}
    \caption{System of decentralised electrical power trading}
    \label{fig:cent_prod_sys}
\end{figure*}
  
  Throughout this paper, the reader will be presented with an overview of relevant building blocks needed for the creation of a decentralised peer-to-peer energy market. 
  First, a discussion about the transition from the status quo towards the new setup will be presented including a discussion of benefits from having in place a decentralised solution.
  This includes the incentives for having a local production and consumption, the new distribution of roles for the actors, the multilayered architecture of such a setup, the need for a decentralised auction house and transaction settling (smart contract) as well as a possible approach for preserving grid stability.
  In addition, an overview of existing projects and initiatives is given.
  
   The remainder of this paper is structured as follows: in the two following sections, the status quo (Sect. \ref{sec:status_quo}) and a possible future power grid architecture with producers, prosumers and consumers (Sect. \ref{sec:architecture}) are discussed. 
  Starting from the innovate aspect of the distributed self-production, Sect.~\ref{sec:roles} explores the incentives and the changes in roles connected with this paradigm shift from top-down to decentralised.
  
   Aspects of local energy trading are then covered in Sect.~\ref{sec:smart_contract_trading}, where a typical \gls{p2p} settlement is explained.
  %Followed in Section \ref{sec:p2p_energy_trading} by an analysis of the local production and local consumption as part of a microgrid grid stability, technological requirements and possible new business approaches. 
  Finally, the chapter is finished by arguing the case for an incomplete energy settlement compensation which will highlight the roles of the network and local distributor operators.
  
  To complement this theoretical development, section~\ref{sec:pilot_and_activities} overviews existing project and activities found in the literature, by analysing some aspects to draw conclusions about the most important themes in the decentralised \gls{p2p} energy market, the \gls{dlt} adopted and the consensus algorithm embraced.
  %The work is concluded by section \ref{sec:conclusion}, where open questions and future research directions are posed.
  %
  Eventually, section \ref{sec:conclusion} recaps this work contribution by drawing some conclusions. 
  Being this a work in progress, many open questions are still on the table and will be addressed in the next stages of the research.

\section{Power Distribution Status quo}\label{sec:status_quo}

% description of current power grid set-up
  To date, the mostly applied electrical power grid topology has implemented a top down approach where the power is star-shaped distributed from the production unit.
  After generating the energy, the power plant transforms it in to a high voltage energy and feeds it into the distribution grid. Over several continuously decreasing voltage levels, the centralised production unit supplies the consumers.
  These actors represent the end nodes of the star-shaped grid.

  Figure \ref{subfig:current_power_top} shows the setup of such a centralised, star-shaped grid architecture, where power is mainly unidirectional supplied from production units to consumers. In such a setting the only allowed action for consumers is to receive electricity from a small number of dedicated producers. 
  
  % definition of grid parties: producer, prosumer, consumer
  In a classic model of the production, distribution and use of electrical energy, a producer feeds a distribution grid that connects the consumers. 
  A \gls{lpd} is the primary point of contact for the consumers and it controls the energy flow.
  Figure \ref{subfig:current_grid_parties} represents typical energy profiles for the actors in the current star-shaped power grid set-up.
  Worth to note is that a typical consumer obtains its full energy need from the grid, while a producer shows a large production amount coupled with a negligible amount of self consumption. 
   
  As can be seen in the figure, there is another group of grid participants, known as prosumers.
  These actors, along with the dominating consumption, also show a small self-generation. 
  In the past, only few prosumers managed to cover parts of their own consumption with sel-generated energy from renewable sources. 
  
  In case of a self-gerneration of the individually produced energy, it simply reduces the prosumer overall consumption from the public grid.
  Thus, improvements and new technologies allow prosumers to obtain a higher proportion of self-sufficiency. 
  Over time, the self production levels have increased significantly, leading to moments of overproduction, during certain times of the day. 
  This overproduction should either be stored in local batteries for a later reuse or be fed into the grid.  
  
  Storage (batteries) however are expensive (both for buying and installing) and their useful life span is limited. The resulting financial viability makes them unappealing for a storage of significant amount~\cite{broering2017simulation}.
  
\begin{figure}[t]
    \centering
    \includegraphics[width=0.48\textwidth]{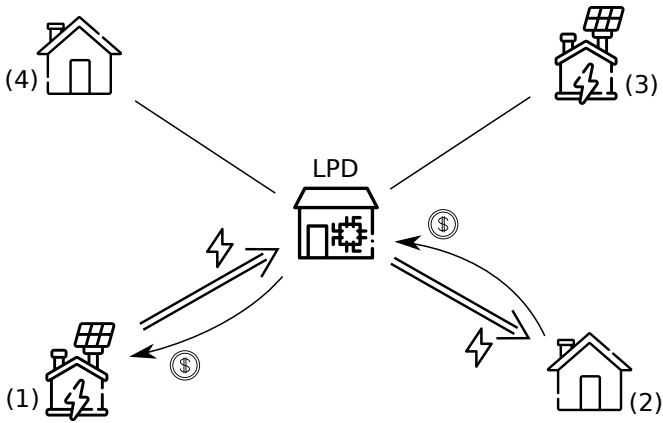}
    \caption{Energy trading in the present power distribution system: it obliges to deal with the \gls{lpd}. A direct trading is not possible.}
    \label{fig:star_energy_cash_flow}
  \end{figure}

  If a prosumer sells his or her own generated power to the \gls{lpd} by feeding it back into the public grid, only a small amount of money is usually paid out. In fact, there is no market that defines the price but only a \gls{lpd} can buy the generated energy, thus maintaining an artificially low price.
  This is because the \gls{lpd} is the only available trading partner, as shown in figure~\ref{fig:star_energy_cash_flow}. 
  Here an excess of self-generated energy in (1) is sold to the \gls{lpd} for a fix price defined by the \gls{lpd} itself. 
  Upon request from (2), \gls{lpd} will provide it for a  higher price.
  The consequence is that prosumers have no incentives to feed in the excess of energy from self generation.
  
  The star-shaped distribution setting dominated the energy market due to the limits of available technologies for small-scale electrical power generation and the significant investments needed, and thus elevated entry barrier, for installing renewable energy production infrastructure. Additionally, the existing legal framework acts as a further barrier.
  
  In fact, the law does not allow buying energy from multiple contractors, even though in some countries it is possible to choose from a limited list of \gls{lpd}es, such as in England~\cite{ofgem2020}.
  
  Hence, there will always be a remaining dependency on an external power distributor in a system that is designed to be hierarchical and top-down. This is particularly problematic in an environment with a continuously increasing decentralised power production.

\section{Architecture of a Peer-to-Peer Energy Market}\label{sec:architecture}

An alternative approach which takes into account an increasing decentralisation of the power production and allows for a bottom-up based system architecture is the Peer-to-Peer energy market.

\begin{figure}[t]
    \centering
        \includegraphics[width=0.47\textwidth]{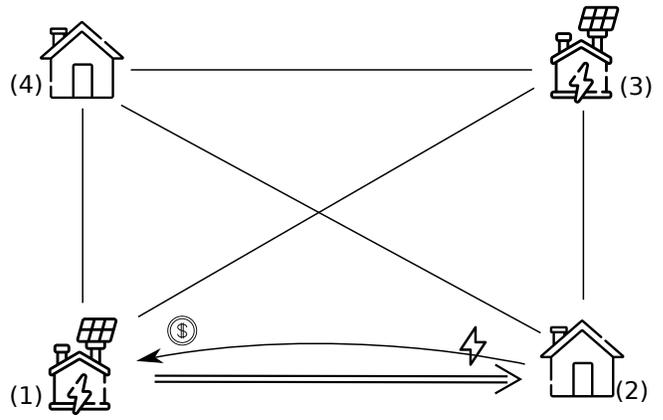}
        \caption{A deregulated system allows for a direct payment and energy exchange between two parties (e.g. prosumers).}
        \label{fig:mesh_energy_cash_flow}
\end{figure}

In such a system, energy trading can happen bilaterally between two actors, a buyer and a seller, without having the \gls{lpd} as an intermediary. Figure~\ref{fig:mesh_energy_cash_flow} displays this direct ordered settlement between the parties (1), providing the electricity, and (2), receiving and financially compensating it. 

\begin{figure*}[t]
    \centering
    \includegraphics[width=0.7\textwidth]{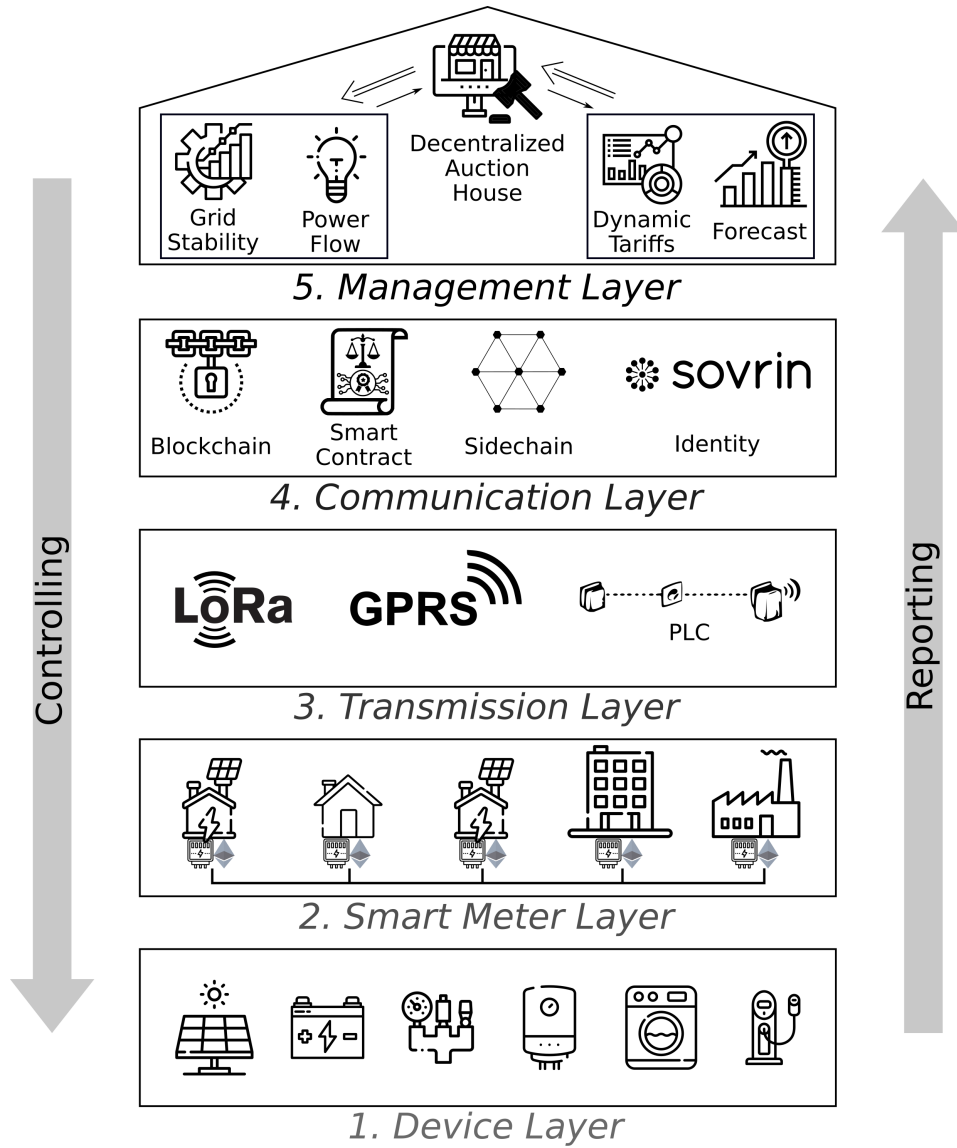}
    \caption{Process architecture for participating on a deregulated energy market system.}
    \label{fig:layered_process}
\end{figure*}{}

For a Peer-to-Peer energy market to work effectively, a multi-layered architecture has to be in place. Figure~\ref{fig:layered_process} provides insights into the different layers, their roles and inter-dependencies. 
The layer are ordered upwards based on the data aggregation level and  structured as follows:

\begin{enumerate}
\item \textbf{Device Layer} includes all physical devices, either present in a household/factory, or distributed in the electric distribution network for stabilisation purposes such as batteries. Data produced in this layer ranges from simple consumption measurement up to advanced information about device usage and status. This is also due to the lifespan of the electrical apparatus, usually spanning one or more decades, and making necessary to actively include also outdated appliances.
%Problem : different data quality/granularity/richness(informativeness) 
The main issues in this level are the different data quality, the granularity of the data available and its richness. 
Consequently, the quality and informativeness of the aggregated information provided to the upper level can be affected. In fact, many of the current devices deployed do not have an on-board communication interface.

\item \textbf{Smart Meter Layer} represents the place where the information of the \textit{Device Layer} are aggregated into a single entity, for being published in the public space (post-meter). 
The Smart Meter can be owned by either the \gls{lpd} or the \gls{dso}. It is imperative that the devices in this tier are not owned by the end users, in order to avoid any risk of self-beneficial manipulation.
In this way, it can work as a separation of concerns between the pre-meter and post-meter part of the electric network. Consequently, all power flows pre-meter is out of control from the \gls{lpd} and \gls{dso} and is transparent to the public network.
The \textit{Smart Meter Layer} can be organised hierarchically, meaning single apartments/units/productive units can have a local meter, but they can be aggregated at a higher level such as buildings/companies or even blocks. 
%Issue: weak HW, issue in coupling with other SW service/sealed, reading intervals (15 mins),  different providers, different communication protocols and standards, roll out is expensive
The first problem observed here is the original scope for which Smart Meters were designed: measuring and pushing data in a certain time interval. No additional  complex functionality was envisioned, thus the very limited specification of the \gls{hw} used, and difficulty in adopting it for usage in a \gls{p2p} market.
An additional limitation stems from their sealed nature that prohibits any manipulation of the on-board software by a third party, making it difficult to adapt it to different requirements, such as "faster" communication intervals of less than 15 minutes.
In a fragmented market, with multiple smart meter providers, a common setting is the presence of different communication protocols and standards. This poses a barrier for implementing new solutions because it is necessary to cope with all of them. 

\begin{figure*}[t]
    \centering
    \begin{tabular}[\textwidth]{c|c}
        \centering
        \multirow{4}{*}{
        \begin{subfigure}[t]{0.65\textwidth}
        \vspace{-1.8cm}
           \includegraphics[width=\textwidth]{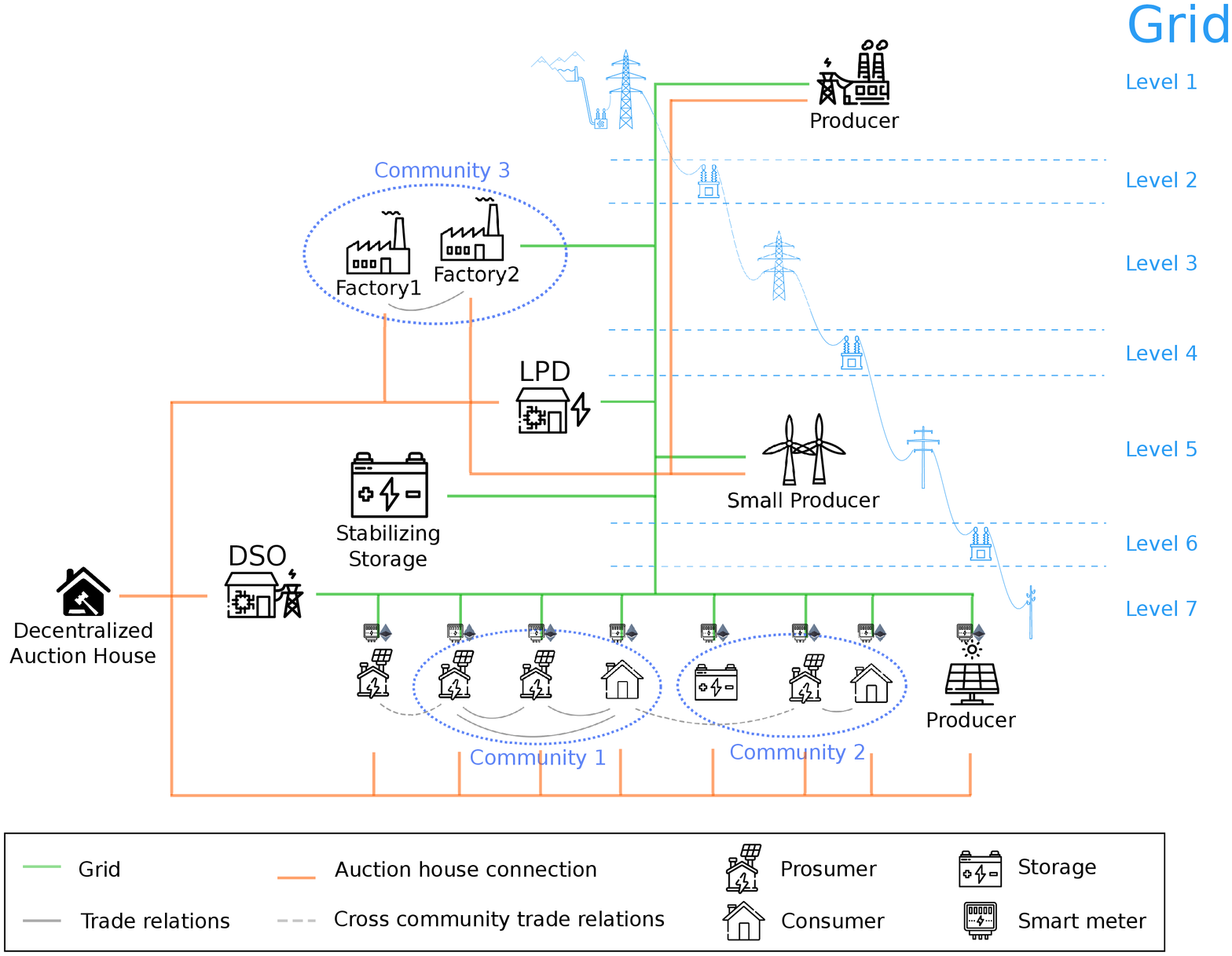}
            \caption{A possible future decentralised power distribution setup, where energy can be exchanged between all parties within a community network. 
            The green links represent the grid, and the orange ones the trading channel.
            Communities are built based on geographical proximity and and willingness to participate with the aim of achieving higher rate of local production and consumption.
            }
            \label{subfig:p2p_power_top}
        \end{subfigure}
        }&
        \begin{subfigure}[]{0.3\textwidth}
                \includegraphics[width=\textwidth]{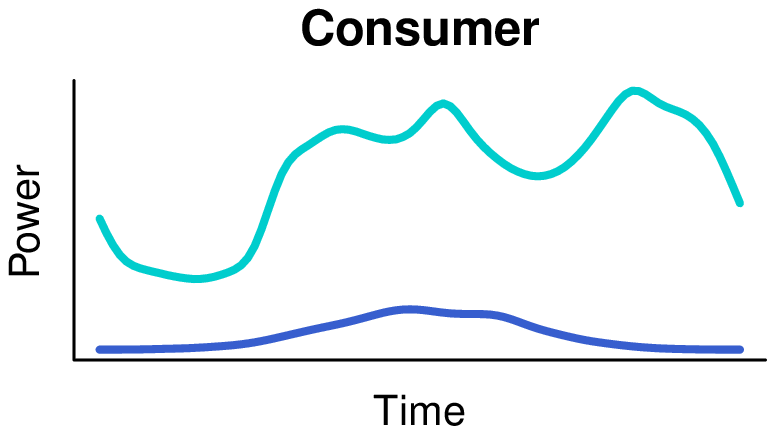}
        \end{subfigure}\\
        &
        \begin{subfigure}[]{0.3\textwidth}
                \includegraphics[width=\textwidth]{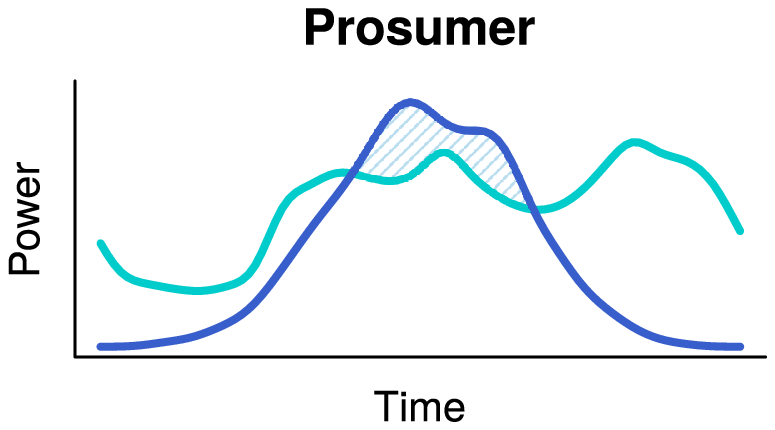}
        \end{subfigure}\\
        &
        \begin{subfigure}[]{0.3\textwidth}
                \includegraphics[width=\textwidth]{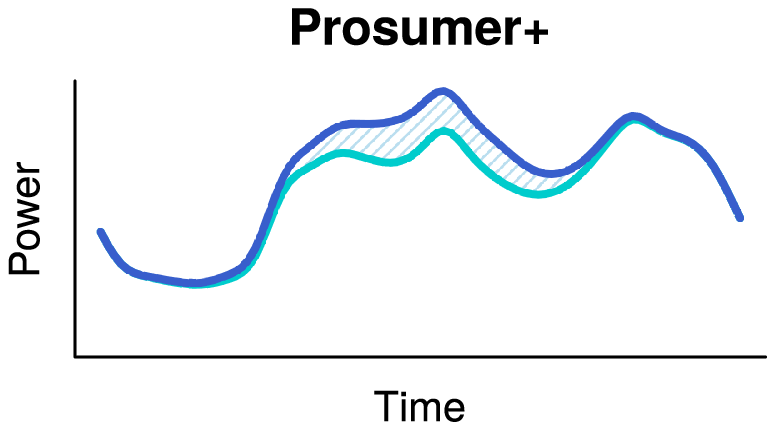}
        \end{subfigure}\\
        &
        \begin{subfigure}[]{0.3\textwidth}
                \includegraphics[width=\textwidth]{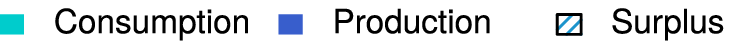}
                \caption{Actors in a possible future mesh power grid set-up. The producer is not represented, having a similar profile to fig.~\ref{subfig:current_grid_parties}.}
                \label{subfig:p2p_grid_parties}
                \vspace{0.2cm}
        \end{subfigure}\\
    \end{tabular}
    \caption{System of decentralised electrical power trading}
    \label{fig:dec_prod_sys}
\end{figure*}

\item \textbf{Transmission Layer} concentrates in the communication channel between Smart Meters and \gls{dso} or \gls{lpd}. It can be implemented using different technologies, such as \gls{lora}, \gls{gprs} and \gls{plc}.
%issues: not always guaranteed connection, data throughput too low (eg: valley), security 
The main issue is the communication reliability: these protocols are not immune from connection drops and offer a limited bandwidth, in particular as consequence of topological and geographical constraints. As a consequence the achievable data throughput can be negatively impacted. 
High security standards should be adopted on top of the used protocol to ensure data reliability and trustability.

\item \textbf{Communication Layer} provides the functionalities for the integration of end users into the digital \gls{p2p} marketplace. This involves the identification and governance of users and \gls{hw}s, which serves as the basis for an automated matching between energy offers and requests. 
Here smart contracts regulate how the energy consumption and production aggregated data is publicly registered in the immutable, distributed ledger, as well as, the commitments for energy providing and consuming.
%issues: choice of protocols, usage of the electricity (PoW vs. others), GDPR
One critical aspect of this layer stems from the fact that most protocols are in an early stage of development and adoption. This makes it difficult to make sustainable long term planning.
The choice of protocol impacts multiple aspects, such as scalability, energetic footprint and compliance with privacy-related regulations.

\item \textbf{Management Layer} is the top level, responsible for processing data coming from lower layers ("Reporting" arrow), taking decisions, propagating them downwards ("Controlling" arrow) and, therewith, regulating the behaviour of the system.
Amongst the main functionalities are energy production and consumption forecasts, regulation of the power flow and grid stability, dynamic tariffs setting and order settlement, as part of the decentralised auction house.
%issues: predictability, gaming of the system, stability of prices, order settlement, grid stability
The software applications sitting in this level serve as oracles, providing knowledge on the networks status and expected behaviours. All these insights are forwarded to the decentralised auction house, which serves as the logical hub of the \gls{p2p} system. Here decisions are taken and then propagated downward throughout the different layers of the multi-layered architecture.
\end{enumerate}

By implementing this architecture, all the basic elements for a full fledged \gls{p2p} energy market are in place. This opens up various opportunities and incentives for building a new market structure, which focuses on decentralisation.

\section{Microgrids, Incentives and Roles}\label{sec:roles}

The transition of the network from a typically top-down approach to distributed production has implication on the exchange structure. The current trading paths are designed to always go over a centralised hub (e.g: \gls{lpd}), which serves as an intermediary with the power of imposing the tariff structure.
The \gls{p2p} energy market, on the other end, follows a different approach, allowing intermediary-free direct trade between two parties, using a decentralised auction house as exchange platform. One of the key features of this auction house is to provide barrier-free egalitarian access to the energy market for all the actors. 
Consequently,  the price is not set anymore by a single, privileged entity, but is decided by the free-market rules, following demand and supply.
The resulting market freedom allows different incentives to be put in place for the end user supporting its maximisation of return.
The excess in self-production can be either saved locally, in battery-based storage, or directly fed-in into the grid, based on the fair-priced possibility of an open market.
Given the need to pay (network transportation fee) for the actual usage of the local grid to the \gls{dso}, the optimal approach is to have only local power exchanges between two actors of a community. 

These abstract entities are represented in fig.~\ref{fig:dec_prod_sys} with dashed blue ovals, and are based on local geographical proximity and willingness to participate. As shown by a recent work, there is a structural limit for the effectiveness of this entities without affecting the global grid stability. It was demonstrated that the threshold is 10 participants, even when of different types~\cite{schopfer_assessment_2019}.
Their objective is to foster a higher level of local production and consumption, and they are known as \textbf{microgrid}, due to their low limit in participants number, with respect to the typical grid dimension.

From a theoretical point of view, each microgrid could aim to a full self-sustainable environment, where local production equals the local consumption. This is also known as self-sustainability or autarchy. Despite the positive impacts such a fact will hold, this is globally a non realistic assumption: in fact exchange can happen also outside the local community, either because the demand and the request are not balanced within, or because some end user is not included in any microgrid, such as the first prosumer on the left of fig.~\ref{fig:dec_prod_sys}.
As the most part of the energy trades are supposed to happen inside the microgrid, the action house responsible for it has also the possibility to implement road pricing for fostering the self-sufficiency.
In this context, road pricing refers to paying exclusively for the costs associated with transferring energy between seller and buyer on the available shortest path. This is supposed to represents only a fraction of the current transmission fees, since the entire transmission grid is not used for this order settlement. 

%% TODO: --> 80\% of the grid cost are on the lower level connection (in CH, find a reference) 

For a better understanding of the new \gls{p2p} energy market, the potential new roles are described in further details in the following section:

\begin{enumerate}
\item \textbf{Consumer} fulfils a similar role as in the status quo. The main difference being that the consumer can now also access the auction house and thus profit from its services.
This includes generating a personalised energy mix (combination of different energy resources), but also paying lower transportation fees, supported by the road pricing incentives model.
Consumer preferences can be either managed manually or by using an artificial agent to automate the transactions. This could potentially allow the consumer to adopt optimisation strategies.

\item \textbf{Prosumer} is a special type of \textit{Consumer}, that presents a certain level of self-production. It can presents moment of surplus in self-production, which is then fed into the grid and financially compensated, based on supply and demand in an open market place. Pre-meter trading allows for a zero-transportation-fee (ZTF) transaction, while post-meter trading inside the boundary of the microgrid can lead to reduced transportation fees.
  
\item \textbf{Prosumer+} is a type of \textit{Prosumer} with an average self-production that lies above the consumption. As such there is an increased need for feeding-in the energy surplus into the market. A distributed \gls{p2p} energy market provides the most significant benefits for this category of market participant. 

\item \textbf{Producer} holds the same role as in the status quo. It can also profit from the decentralised \gls{p2p} system by being able to directly sale to the \textit{Consumer} without a middle-man. Additionally, a significant leverage to affect the price is given, in consequence to the large availability of production and storage capacity.
  
\item \textbf{DSO} ensures a congestion-less energy transmission in a free energy market by providing the grid and all smart meters. It receives a fee on every transaction for these services (road pricing).
  
\item \textbf{LPD} is responsible for the stability of the grid as a service for the community. Using large storage capabilities, it can intervene for stabilising the unbalances in the offer/demand, or to backup commitments that have not be honoured by one of the party. These storage facilities do not usually participate in the open market. Consequently, these services are subject to a stability fee, payed by the party that breaches the contract.  

\end{enumerate} 

\begin{figure}[b]
    \centering
    \includegraphics[width=0.46\textwidth]{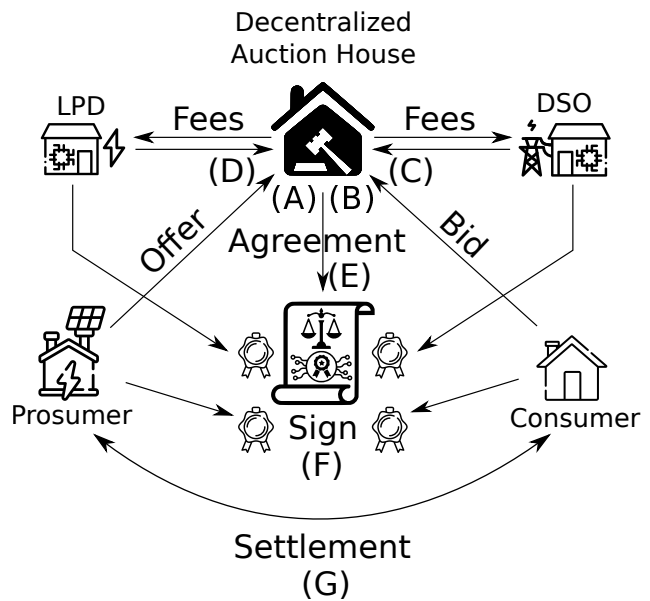}
    \caption{The trading process, centred around the Decentralised Auction House.}
    \label{fig:trading_procedure}
\end{figure}{}  

\begin{figure*}[t]
    %\todo{try to get the original figure from deke -> unable to convert figure to vector graphic.}
    \centering
    \includegraphics[width=0.9\textwidth]{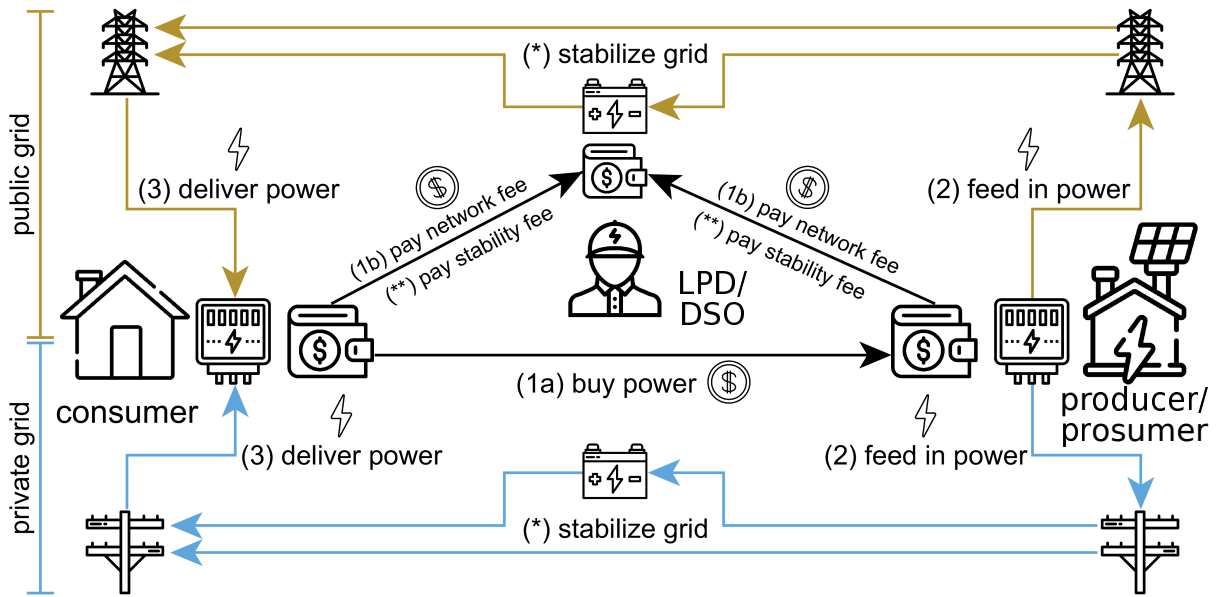}
    \caption{Overview of a peer-to-peer network in a smart grid application \cite{deke_offchain_2020}. All members (producer, prosumer, consumer) participate on the same level while a 3rd party utility company provides the grid and all smart meters.}
    \label{fig:overview_p2p_in_smartgrid}
\end{figure*}{}

The presence of so many different roles and conflicting interests has a significant impact on the trading architecture. 
As such it is critical to introduce a well-thought design that ensure that none of the parties has an advantageous position to force the system into an unstable state.

\section{Smart Contract Trading}\label{sec:smart_contract_trading}

Figure~\ref{fig:trading_procedure} shows a potential Smart Contract architecture for a decentralised  \gls{p2p} energy network. 
The previously described parties cooperate to create a settlement between a prosumer and a consumer.
In order to achieve this, the following non-consecutive steps should be included:

\begin{enumerate}
    \item[(A)] the Prosumer offer a given amount of self-produced available energy for sale on the network, at a certain given price range. The offer include the energy origin. The auction house generates energy tokens in accordance with the declared energy amount that can be provided into the grid in the future. These (utility) tokens are made available within the digital wallet of the Prosumer.
    \item[(B)] A consumer expresses the desire to purchase energy in the form of tokens in the near future. This request includes quantity, price range and type of energy, which is part of the personalised energy mix.   
    \item[(C)] Once a matching between prices, energy origin and quantities has occurred, the shortest path between the two parties on the grid is calculated by \gls{dso} and the transportation fees derived. 
    \item[(D)] In addition the \gls{lpd} computes the fees for stabilise the grid. These fees, in form of a collateral, are required to cover the cost of a potential intervention, in case one of the parties does not honour the agreement.
    \item[(E)] The full smart contract is generated by the decentralised Auction House, including all the terms and conditions.
    \item[(F)] Multi-signature is added to the Smart Contract, in order to enforce its validity. 
    \item[(G)] The tokens held by the Prosumer and promised to the Consumer are burned contextually to the release of the energy. From the Consumer side, the financial compensation for the energy is released and transferred to the Prosumer digital wallet. This constitutes the Settlement of the Smart Contract.
\end{enumerate}

If the settlement fails, either for a lack of release of the promised energy from the Prosumer or the inability for the Consumer to absorb the demanded energy, a fail-safe system needs to kick-in. This is necessary to avoid significant fluctuations in the voltage or frequency, which can damage powered devices.
In the following section such a system is described in a more detailed way.
To date, smaller and flexible producers as pumped-storage power plants compensate the divergence between demand and offer, while slow and large power plants are used for a basis load.

Figure~\ref{fig:overview_p2p_in_smartgrid} shows the compensation of an incomplete settlement. The \gls{dso} jumps in by either absorbing the excess energy in the grid (Consumer failure to absorb the energy)  or by providing the lacking amount of power (Prosumer impossibility to deliver the agreed amount). For fairness of the network, the party that correctly honours the smart contract should not be affected by the failure to comply with it on the other side. This means that the collateral deposited (stability fees) of the non-compliant partner only should be used for compensate the stabilisation task.
Also in the private part of the network (pre-meter) can exist stabilisation facilities, in the form of privately owned batteries. These are responsible to provide a stability in the internal network.

Despite all the open issues presented here, a significant amount of research is already available in the area of \gls{p2p} distributed energy market. 
A set of activities and pilot installations exists covering one or more of the criticalities presented in this paper. 
The next section will present a reasoned review of some of the existing studies.

\section{Activities and Pilot installations of P2P Energy Markets}\label{sec:pilot_and_activities}

Starting from some recent reviews about the \gls{blc} technologies usage and perspectives for the energy domain~\cite{andoni_blockchain_2019, johanning_blockchain-based_2019, wang_when_2019}, a set of 42 project related to \gls{p2p} were identified. 

% The analysed dimensions include the status (active or not), the type (an autonomous project or an effort from an existing company, already on the market), the availability of a public document (scientific or white paper) detailing the project approach and outcomes, the reference country, the focus on the smart grid aspect, the scope (local, regional or national/global), the year of activities start, ...
% For all these dimensions, the data are reported when known/available and left empty if unknown.

Five main aspects are considered in the analysis, namely, 
the (I) country where the activity is rooted, 
the (II) main focus of the project, 
the (III) geographic scope, 
the (IV) \gls{blc} technology and 
the (V) category of consensus algorithm adopted, currently (V.a) and in the future (V.b).

\begin{figure}[t]
    \centering
    \resizebox{0.95\columnwidth}{!}{
        \begin{tikzpicture} 
            \begin{axis}[
            %title    = Frequency of initiative for country,
            xbar,
            ytick             = data,
            y axis line style = { opacity = 0 },
            axis x line       = none,
            tickwidth         = 0pt,
            %enlarge y limits  = 0.2,
            enlarge x limits  = 0.2,
            nodes near coords,
             symbolic y coords = {
             Others, 
             Singapore, Japan, Belgium, Australia, UK, Switzerland, France, Germany, US, Netherlands,
             %Netherlands, US, Germany, France, Switzerland, UK, Australia, Belgium, Japan, Singapore, 
             %Brazil, China, Denmark, India, Israel, Italy, Lithuania, New Zealand, Romania, Slovenia, Spain, Thailand, 
             %Others
             },
            ]
            \addplot coordinates { (7,Netherlands) (6,US) (4,Germany) (3,France) (3,Switzerland) (3,UK) (2,Australia) (2,Belgium) (2,Japan) (2,Singapore) (9,Others) };
            \end{axis}
        \end{tikzpicture}
    }
    \caption{Expected evolution of the different families of consensus algorithms, in percentage on the analysed projects.}
    \label{fig:p2p_countries_distribution}
\end{figure}
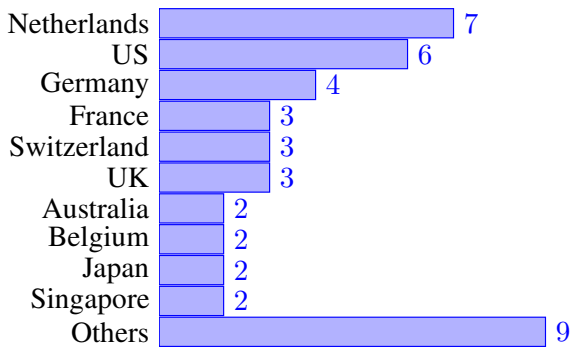

For an initial classification of the countries where the activities originated, figure~\ref{fig:p2p_countries_distribution} provides a cumulative view. Due to space reasons, every country with just a single project is collected into the \textit{Others} group. A clear interest in some European country is evident (in particular, the Netherlands forsaw this as one of the solutions towards a completely gas-free energy production). Also the United States of America and the centre of the EU (Germany and France) have a significant number of activities. Switzerland and UK have 3 reported projects each, where Australia, Belgium, Japan and Singapore present 2 entries each.

Table~\ref{table:kind_of_system} presents a division of this set based on the (II) type of application (main focus) of the project itself. 
The \emph{Smart Grid} category groups projects where the attention is either on providing a P2P network detached from the traditional star-shaped energy distribution or on designing the full architecture and the relevant assets for creating such a system. On contrast, \emph{\gls{p2p} Platform} represents activities that focuses on the energy trading platforms without an explicit connection to the energy measurement and the relevant oracles used to providing information to the blockchain.
Can be noted that the majority of them focus around smart grid, whether about a quarter main objective is in \gls{p2p} platform support. The remaining ones aim to different topics and is here collected under the \emph{Other} class.

\begin{figure}[t]
    \centering
    \resizebox{0.75\columnwidth}{!}{
        \begin{tikzpicture}
            \pie[
            %cloud, 
            %text=inside, 
            %scale font,
            %pos={4,16}, 
            %explode={0.2,0.2,0.2,0.2,0.2,0.2,0.2},
            rotate = 90,
            color = {red!14, red!29, red!43, red!58, red!72, red!87, red!100}]{
                    64/Ethereum,
                    9/Tendermint,
                    9/Proprietary,
                    4.5/Multichain,
                    4.5/Quasar,
                    4.5/Sky-Ledger,
                    4.5/Stellar
                    }
        \end{tikzpicture}
    }
    \caption{Distribution of applied \gls{dlt} in P2P energy trading projects.}
    \label{fig:p2p_blockchains}
\end{figure}
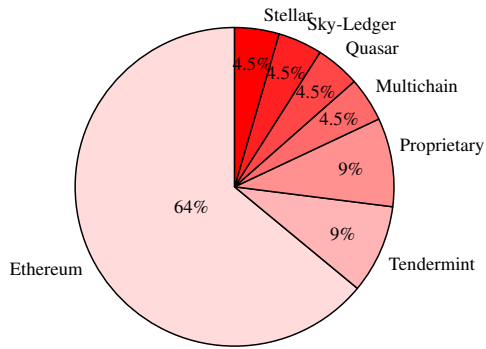 

Regarding the geographical scope of the activities (III) , table~\ref{table:system_geographical_scope} reports the division into local, regional and national/global. 
Here is defined as local an activity that is limited to a small number of selected participants, located in close vicinity, such as for a neighbourhood or a small city district. These are typically small P2P communities especially formed for energy trading purposes. For the regional level, is taken the typical area of coverage of a \gls{lpd}, such as cities and metropolitan areas, whether the national/global level covers multiple regional (or local) scopes. 
As potentially noted, the local scope is the prevalent  focus for half of the reported exercises. This is also in accordance with the results for the application, as the smart grid focus is usually correlated with a local target, for easiness of introduction and to avoid conflicts with the current legal framework in the energy market.
In fact 12 of of 16 projects with focus on smart grid have also a local deployment scope. 
Another interesting aspect is the prevalence of globally scoped activities over the regional ones, likely due to the broader expected impact of the project, whenever the regulating framework should anyway be taken into account.

  \vspace{0.3cm}
  \begin{minipage}[b]{.20\textwidth}
    \centering
      \captionof{table}{Geographical scope (dimension) of the systems}
      \begin{tabular}{l r}
           \hline
           \multicolumn{2}{c}{Activity Dimension} \\
           \hline
           Local & 50\% \\
           Regional & 9\% \\
           National & 41\% \\
           \hline
      \end{tabular}{}
      \label{table:system_geographical_scope}
  \end{minipage}\qquad
  \begin{minipage}[b]{.20\textwidth}
  \centering
      \captionof{table}{Applications, that the systems are used for}
      \begin{tabular}{l r}
            \hline
            \multicolumn{2}{c}{Provided System} \\
            \hline
            Smart Grid & 52\% \\
            P2P Platform & 24\% \\
            Other & 24\% \\
            \hline
      \end{tabular}{}
      \label{table:kind_of_system}
  \end{minipage}
  \vspace{0.3cm}
  
The next relevant aspect (IV) scrutinises which \gls{dlt} is adopted for the project.       
As evident from figure~\ref{fig:p2p_blockchains}, there a strong predominance of the \emph{Ethereum} technology. 
A non-conclusive list of factor that can explain this phenomenon exists. Considering the relatively young age of this business field, the first players entering the domain are genrally perceived as most trustable and paving the path. In fact Ethereum got a significant traction in the early stage of \gls{dlt} adoption, also because it presents very good documentation and a significant amount of well-designed and comprehensible examples for the most diffused functionalities.
Cascade, whose adoption creates a vibrant and active community around the software, which guarantee continuous updates and easier access to programmable ready use for the underlying protocol. This is also an implicit signal that adopting Ethereum will be less risky from the business point of view<< as this interest will realistically support the assumption that the technology will be still in place and usable in a 5-years horizon.
One definitive aspect that oriented the adoption towards Ethereum is the fact that the protocol natively supports the IRC token standard, making it very easy to generate the type of utility token needed based on the specific asset that should be covered. All the Bitcoin family does not offer natively such a functionality.

This demonstrates that the native possibility of using smart contacts for the energy transactions is an important task.
Other initiatives adopted the new concept of \emph{Multichain} (an open infrastructure, where the different \gls{dlt} solutions can coexist, also with the possibility of exchanging currency and tokens amongst them) for allowing a smooth potential integration of already existing local initiatives in the distributed \gls{p2p} energy market. 

\begin{figure}[t]
    \centering
    \resizebox{0.75\columnwidth}{!}{
        \begin{tikzpicture}   
            \pie[
            rotate = 90,
            %pos={0,0}, 
            %explode={0.2,0.2,0.2,0.2,0.2,0,0.2,0.2,0.2},
            color = {orange!11.1, orange!22.2, orange!33.3, orange!44.4, orange!55.5,
            orange!66.6, orange!77.7, orange!88.8, orange!100}]{
                    55/PoS,
                    10/PoA,
                    5/AuRa,
                    5/PoW,
                    5/Obelisk,
                    5/PBFT,
                    5/PoC,
                    5/PoST,
                    5/TCP}
        % \end{tikzpicture}
        \end{tikzpicture}
    }
    \caption{Distribution of adopted type of consensus algorithms for \gls{p2p} blockchain applications: future planned algorithms.}
    \label{fig:p2p_consalg_future}
\end{figure}
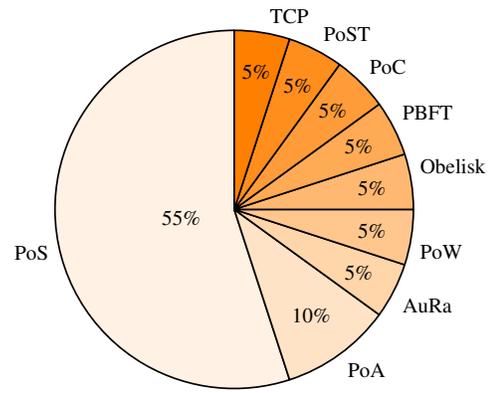

The fourth aspect took into consideration is probably the most critical issue up to date for the adoption of \gls{dlt} solutions in energy market, and has to do with scalability and energy consumption for running the system. This feature is the consensus algorithm adopted. Figure~\ref{fig:p2p_consalg_today} and figure~\ref{fig:p2p_consalg_future} present respectively the current status and the future expected consensus approach that the analysed activities declare.

\begin{figure}[t]
    \centering
    \resizebox{0.75\columnwidth}{!}{
        \begin{tikzpicture}            
            \pie[
            rotate = 90,
            %pos={0,8}, 
            %explode={0.2,0.2,0.2,0.2,0.2,0.2,0.2,0,0.2},
            color = {lime!11.1, lime!22.2, lime!33.3, lime!44.4, lime!55.5,
            lime!66.6, lime!77.7, lime!88.8, lime!100}]{
                    50/PoW,
                    10/PoA,
                    10/PoS,
                    5/AuRa,
                    5/Obelisk,
                    5/PBFT,
                    5/PoC,
                    5/PoST,
                    5/TCP}
        \end{tikzpicture}
    }
    \caption{Distribution of adopted type of consensus algorithms for \gls{p2p} blockchain applications: currently applied algorithms.}
        \label{fig:p2p_consalg_today}
\end{figure}
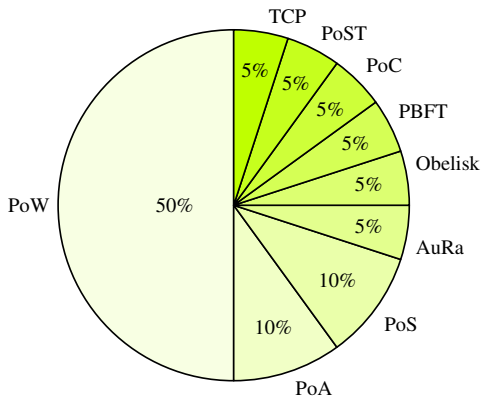    

It is noteworthy that in absence of information regarding these aspects in the documentation or publication from the project itself, we assumed that the "native" agreement approach from the chosen \gls{dlt} is preserved. This analysis is not run at the level of the specific algorithm, but aggregating them based on the main underlying functioning mechanism. This is also useful to draw some general conclusions about the limitation and the offered properties.
What can be noted here is the moving from a predominance of computationally intensive and energy voracious approaches towards more scalable algorithms that stress the recognition of nodes commitment to serve the \gls{dlt}, in term of resources specifically and uniquely devoted to it. 
The current status, in fig.~\ref{fig:p2p_consalg_today}, demonstrates the prevalence of \gls{pow} approaches. Here the addition of a new block to the chain involves the resolution of a cryptographic puzzle, operation usually referred to as mining.

% pow vs pos
\begin{figure}[b]
    \centering
    \resizebox{0.95\columnwidth}{!}{
        \begin{tikzpicture} 
            \begin{axis}[
            %title    = Comparison of consensus algorithm classes in percentage,
            xbar,
            y axis line style = { opacity = 50 },
            %axis x line       = none,
            tickwidth         = 0pt,
            enlarge y limits  = 0.2,
            enlarge x limits  = 0.2,
            nodes near coords,
             symbolic y coords = {\gls{pow}, \gls{pos}, \gls{poa}, Others},
            ]
            \addplot coordinates { (50,\gls{pow}) (10,\gls{pos}) (10,\gls{poa}) (30,Others) };
            \addplot coordinates { (10,\gls{pow}) (55,\gls{pos}) (10,\gls{poa}) (25,Others) };
            \legend{Current, Planned}
                   
            \end{axis}
        \end{tikzpicture}
    }
    \caption{Expected evolution of the different families of consensus algorithms ,in percentage on the analysed projects.}
    \label{fig:p2p_consalg_evolution}
\end{figure}
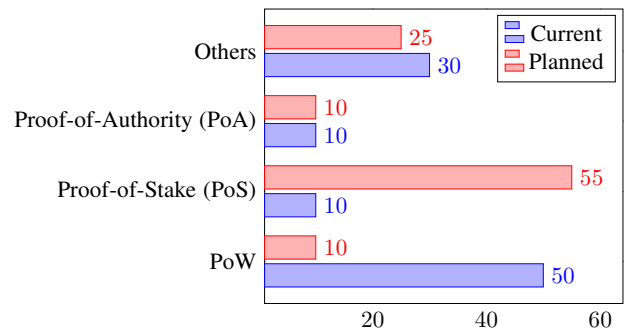 

In contrast fig.\ref{fig:p2p_consalg_future} indicates that other approaches will be privileged in the future, in particular the family of \gls{pos}. Another notable aspect is the fact that proprietary or peculiar algorithms that are adopted for specific reasons, tends to stay in place along the lifespan of the project.
As a final note, the predefined consensus approach of Ethereum is moving in the same direction, due to the request for a significant reduction of its energy consumption\cite{fairley2018ethereum}.
This fact can be clearly read into the aggregated data from figure~\ref{fig:p2p_consalg_evolution}, where the relative frequency of the current and the future type of consensus mechanisms are compared.
By looking only at the \gls{pow} and the \gls{pos} categories, this trend can be clearly seen, with the first reducing of 40\% and the second increasing of 45\%. This can partially be explained by the marketing willingness to present the project as interested into a broader profile of sustainability and scalability, but also by the new Ethereum 2.0, that will move the predefined consensus algorithm to a \gls{pos} solutions.

\section{Conclusions}\label{sec:conclusion}
This work presented an analysis of the energy market, with respect to an increasingly percentage of distribute self-generation of electricity, mainly using small-scale plant, such as \gls{pv} roof-based installations.
It started by analysing the current setting of the energy distribution network and the roles existing in this highly regulated market. It continues by exploring which aspects of the status quo are hindering a real diffusion of a \gls{p2p} decentralised energy market, without privileged positions. By doing this, it also introduces the new roles of Prosumers and defines how the traditional players are affected.
It proposes a model for the layered architecture such a system would require by depicting the features expected in each layer and by exposing the limitation currently existing, due to \gls{hw}, technical or legal aspects of the level.
It continues presenting the current trends towards a more liberalised market, introducing the concepts of Microgrids and their role in facilitating localised exchange between end users, towards a focused higher energetic self-sufficiency.
An analyses of the role evolution in this new \gls{p2p} decentralised setting is provided, stressing the centrality of the \gls{dso} and \gls{lpd} for the energy network financing (through transmission fees collection) and the grid stability (by compensating energy availability oscillations), respectively. 
To showcase how a decentralised market can work, this work present a theoretical model for the use of Smart Contracts in energy trading, stressing the necessary steps for its generation, and showing the mechanisms in place to compensate the network utilities (\gls{dso} and \gls{lpd}) for the services provided to the community of energy users.
Eventually, an analysis of existing activities and pilot projects towards a \gls{p2p} decentralised energy market is reported, showing the prevalent aim, the geographical scope, the type of \gls{dlt} approach and the consensus mechanism suggested. 
With this work, we would like to raise the awareness of the steps still to be done for a real transition towards a decentralised system without privileged actors, but would like to reason with the community interested about models and partial components that can bring us closer to a more fair and ecological-friendly solution for the electricity market of the future.

% use section* for acknowledgment
\section*{Acknowledgment}
%The research leading to this work was partially financed by the KTI/Innosuisse Swiss federal agency, through a competitive call. The financed project KTI-Nr. 27104.1 is called \textit{CVCube: digitale Aus- und Weiterbildungsberatung mittels Bildungsgraphen}. 
The work leading to this paper is partially financed by the European Commission, thanks to the EU Blockchain Observatory \& Forum. This is an initiative sponsored by the European Commission, Directorate-General of Communications Networks, Content \& Technology.
The authors would like to thank all the participants to the activity, in particular the attendees on the online Workshop "Energy and sustainability" for the fruitful discussions and the feedback integrated in this publication.
A special thanks goes to Benjamin Haymond, for his very helpful and precise revision and language editing support of this manuscript.

\bibliographystyle{apalike}
\bibliography{main}

%\listoftodos

\end{document}